\documentclass{PoS}
\usepackage{ifthen} 
\newboolean{articletitles}
\setboolean{articletitles}{true} 
\newboolean{uprightparticles}
\setboolean{uprightparticles}{false} 
\newboolean{inbibliography}
\setboolean{inbibliography}{false} 
\usepackage{xspace} 
\usepackage{upgreek}

\def\MagUp {\mbox{\em Mag\kern -0.05em Up}\xspace}

\ifthenelse{\boolean{uprightparticles}}%
{
 
 \def\Pgamma      {\ensuremath{\upgamma}\xspace}

 \def\Peta        {\ensuremath{\upeta}\xspace}

 \def\Pmu         {\ensuremath{\upmu}\xspace}                 
 \def\Pnu         {\ensuremath{\upnu}\xspace}                 
                  
 \def\Ppi         {\ensuremath{\uppi}\xspace}

 \def\Ptau        {\ensuremath{\uptau}\xspace}

 \def\Pchi        {\ensuremath{\upchi}\xspace}                 
 \def\Ppsi        {\ensuremath{\uppsi}\xspace}

 \def\PDelta      {\ensuremath{\Delta}\xspace}                 
 \def\PXi      {\ensuremath{\Xi}\xspace}                 
 \def\PLambda      {\ensuremath{\Lambda}\xspace}                 
 \def\PSigma      {\ensuremath{\Sigma}\xspace}                 
 \def\POmega      {\ensuremath{\Omega}\xspace}                 
 \def\PUpsilon      {\ensuremath{\Upsilon}\xspace}                 
 
 \def\PB      {\ensuremath{\mathrm{B}}\xspace}                 
                  
 \def\PD      {\ensuremath{\mathrm{D}}\xspace}

 \def\PH      {\ensuremath{\mathrm{H}}\xspace}                 
                  
 \def\PJ      {\ensuremath{\mathrm{J}}\xspace}                 
 \def\PK      {\ensuremath{\mathrm{K}}\xspace}

 \def\PW      {\ensuremath{\mathrm{W}}\xspace}

 \def\PZ      {\ensuremath{\mathrm{Z}}\xspace}                 
                  
 \def\Pb      {\ensuremath{\mathrm{b}}\xspace}                 
 \def\Pc      {\ensuremath{\mathrm{c}}\xspace}                 
 \def\Pd      {\ensuremath{\mathrm{d}}\xspace}                 
 \def\Pe      {\ensuremath{\mathrm{e}}\xspace}

 \def\Ph      {\ensuremath{\mathrm{h}}\xspace}                 
 \def\Pi      {\ensuremath{\mathrm{i}}\xspace}

 \def\Pn      {\ensuremath{\mathrm{n}}\xspace}                 
                  
 \def\Pp      {\ensuremath{\mathrm{p}}\xspace}                 
 \def\Pq      {\ensuremath{\mathrm{q}}\xspace}                 
                  
 \def\Ps      {\ensuremath{\mathrm{s}}\xspace}                 
 \def\Pt      {\ensuremath{\mathrm{t}}\xspace}                 
 \def\Pu      {\ensuremath{\mathrm{u}}\xspace}

}
{
 
 \def\Pgamma      {\ensuremath{\gamma}\xspace}

 \def\Peta        {\ensuremath{\eta}\xspace}

 \def\Pmu         {\ensuremath{\mu}\xspace}                 
 \def\Pnu         {\ensuremath{\nu}\xspace}                 
                  
 \def\Ppi         {\ensuremath{\pi}\xspace}

 \def\Ptau        {\ensuremath{\tau}\xspace}

 \def\Pchi        {\ensuremath{\chi}\xspace}                 
 \def\Ppsi        {\ensuremath{\psi}\xspace}                 
                  
 \mathchardef\PDelta="7101
 \mathchardef\PXi="7104
 \mathchardef\PLambda="7103
 \mathchardef\PSigma="7106
 \mathchardef\POmega="710A
 \mathchardef\PUpsilon="7107
                  
 \def\PB      {\ensuremath{B}\xspace}                 
                  
 \def\PD      {\ensuremath{D}\xspace}

 \def\PH      {\ensuremath{H}\xspace}                 
                  
 \def\PJ      {\ensuremath{J}\xspace}                 
 \def\PK      {\ensuremath{K}\xspace}

 \def\PW      {\ensuremath{W}\xspace}

 \def\PZ      {\ensuremath{Z}\xspace}                 
                  
 \def\Pb      {\ensuremath{b}\xspace}                 
 \def\Pc      {\ensuremath{c}\xspace}                 
 \def\Pd      {\ensuremath{d}\xspace}                 
 \def\Pe      {\ensuremath{e}\xspace}

 \def\Ph      {\ensuremath{h}\xspace}                 
 \def\Pi      {\ensuremath{i}\xspace}

 \def\Pn      {\ensuremath{n}\xspace}                 
                  
 \def\Pp      {\ensuremath{p}\xspace}                 
 \def\Pq      {\ensuremath{q}\xspace}                 
                  
 \def\Ps      {\ensuremath{s}\xspace}                 
 \def\Pt      {\ensuremath{t}\xspace}                 
 \def\Pu      {\ensuremath{u}\xspace}

}

\makeatletter
\ifcase \@ptsize \relax
  \newcommand{\miniscule}{\@setfontsize\miniscule{4}{5}}
\or
  \newcommand{\miniscule}{\@setfontsize\miniscule{5}{6}}
\or
  \newcommand{\miniscule}{\@setfontsize\miniscule{5}{6}}
\fi
\makeatother

\DeclareRobustCommand{\optbar}[1]{\shortstack{{\miniscule (\rule[.5ex]{1.25em}{.18mm})}
  \\ [-.7ex] $#1$}}

\def\en         {{\ensuremath{\Pe^-}}\xspace}   

\def\epem       {{\ensuremath{\Pe^+\Pe^-}}\xspace}

\def\mup        {{\ensuremath{\Pmu^+}}\xspace}
\def\mun        {{\ensuremath{\Pmu^-}}\xspace} 

\def\neu        {{\ensuremath{\Pnu}}\xspace}
\def\neub       {{\ensuremath{\overline{\Pnu}}}\xspace}
\def\neue       {{\ensuremath{\neu_e}}\xspace}
\def\neueb      {{\ensuremath{\neub_e}}\xspace}
\def\neum       {{\ensuremath{\neu_\mu}}\xspace}
\def\neumb      {{\ensuremath{\neub_\mu}}\xspace}
\def\neut       {{\ensuremath{\neu_\tau}}\xspace}
\def\neutb      {{\ensuremath{\neub_\tau}}\xspace}
\def\neul       {{\ensuremath{\neu_\ell}}\xspace}
\def\neulb      {{\ensuremath{\neub_\ell}}\xspace}

\def\g      {{\ensuremath{\Pgamma}}\xspace}

\def\dquark    {{\ensuremath{\Pd}}\xspace}

\def\squark    {{\ensuremath{\Ps}}\xspace}

\def\cquark    {{\ensuremath{\Pc}}\xspace}

\def\bquark    {{\ensuremath{\Pb}}\xspace}

\def\pion   {{\ensuremath{\Ppi}}\xspace}

\def\pip    {{\ensuremath{\pion^+}}\xspace}
\def\pim    {{\ensuremath{\pion^-}}\xspace}

\def\kaon    {{\ensuremath{\PK}}\xspace}
  \def\Kbar    {{\kern 0.2em\overline{\kern -0.2em \PK}{}}\xspace}

\def\KorKbar    {\kern 0.18em\optbar{\kern -0.18em K}{}\xspace}
\def\Kz      {{\ensuremath{\kaon^0}}\xspace}
\def\Kzb     {{\ensuremath{\Kbar{}^0}}\xspace}
\def\Kp      {{\ensuremath{\kaon^+}}\xspace}
\def\Km      {{\ensuremath{\kaon^-}}\xspace}

\def\KS      {{\ensuremath{\kaon^0_{\mathrm{ \scriptscriptstyle S}}}}\xspace}

\def\Kstarz  {{\ensuremath{\kaon^{*0}}}\xspace}
\def\Kstarzb {{\ensuremath{\Kbar{}^{*0}}}\xspace}

  \def\Dbar    {{\kern 0.2em\overline{\kern -0.2em \PD}{}}\xspace}
\def\D       {{\ensuremath{\PD}}\xspace}

\def\DorDbar    {\kern 0.18em\optbar{\kern -0.18em D}{}\xspace}
\def\Dz      {{\ensuremath{\D^0}}\xspace}
\def\Dzb     {{\ensuremath{\Dbar{}^0}}\xspace}
\def\Dp      {{\ensuremath{\D^+}}\xspace}
\def\Dm      {{\ensuremath{\D^-}}\xspace}

\def\B       {{\ensuremath{\PB}}\xspace}
\def\Bbar    {{\ensuremath{\kern 0.18em\overline{\kern -0.18em \PB}{}}}\xspace}

\def\BorBbar    {\kern 0.18em\optbar{\kern -0.18em B}{}\xspace}

\def\Bu      {{\ensuremath{\B^+}}\xspace}
\def\Bub     {{\ensuremath{\B^-}}\xspace}
\def\Bp      {{\ensuremath{\Bu}}\xspace}

\def\Bd      {{\ensuremath{\B^0}}\xspace}
\def\Bs      {{\ensuremath{\B^0_\squark}}\xspace}
\def\Bsb     {{\ensuremath{\Bbar{}^0_\squark}}\xspace}
\def\Bdb     {{\ensuremath{\Bbar{}^0}}\xspace}
\def\Bc      {{\ensuremath{\B_\cquark^+}}\xspace}

\def\jpsi     {{\ensuremath{{\PJ\mskip -3mu/\mskip -2mu\Ppsi\mskip 2mu}}}\xspace}

  \def\Y#1S{\ensuremath{\PUpsilon{(#1S)}}\xspace}

\def\proton      {{\ensuremath{\Pp}}\xspace}

\def\neutron     {{\ensuremath{\Pn}}\xspace}

\def\Deltares    {{\ensuremath{\PDelta}}\xspace}

\def\Xires       {{\ensuremath{\PXi}}\xspace}

\def\Lz          {{\ensuremath{\PLambda}}\xspace}
\def\Lbar        {{\ensuremath{\kern 0.1em\overline{\kern -0.1em\PLambda}}}\xspace}
\def\LorLbar    {\kern 0.18em\optbar{\kern -0.18em \PLambda}{}\xspace}
\def\Lambdares   {{\ensuremath{\PLambda}}\xspace}

\def\Sigmares    {{\ensuremath{\PSigma}}\xspace}

\def\Omegares    {{\ensuremath{\POmega}}\xspace}

\def\Lb      {{\ensuremath{\Lz^0_\bquark}}\xspace}

\newcommand{\decay}[2]{\ensuremath{#1\!\to #2}\xspace}         

\def\to                 {\ensuremath{\rightarrow}\xspace}

\def\CP                {{\ensuremath{C\!P}}\xspace}

\newcommand{\DGs}{{\ensuremath{\Delta\Gamma_{\squark}}}\xspace}

\newcommand{\phid}{{\ensuremath{\phi_{\dquark}}}\xspace}

\newcommand{\phis}{{\ensuremath{\phi_{\squark}}}\xspace}

\newcommand{\etag}{{\ensuremath{\varepsilon_{\mathrm{tag}}}}\xspace}

\def\AT#1     {\ensuremath{A_{\mathrm{T}}^{#1}}\xspace}           

\def\C#1      {\ensuremath{\mathcal{C}_{#1}}\xspace}                       
\def\Cp#1     {\ensuremath{\mathcal{C}_{#1}^{'}}\xspace}                    
\def\Ceff#1   {\ensuremath{\mathcal{C}_{#1}^{\mathrm{(eff)}}}\xspace}        
\def\Cpeff#1  {\ensuremath{\mathcal{C}_{#1}^{'\mathrm{(eff)}}}\xspace}       
\def\Ope#1    {\ensuremath{\mathcal{O}_{#1}}\xspace}                       
\def\Opep#1   {\ensuremath{\mathcal{O}_{#1}^{'}}\xspace}                    

\newcommand{\unit}[1]{\ensuremath{\mathrm{ \,#1}}\xspace}          

\newcommand{\tev}{\ifthenelse{\boolean{inbibliography}}{\ensuremath{~T\kern -0.05em eV}\xspace}{\ensuremath{\mathrm{\,Te\kern -0.1em V}}}\xspace}
\newcommand{\gev}{\ensuremath{\mathrm{\,Ge\kern -0.1em V}}\xspace}
\newcommand{\mev}{\ensuremath{\mathrm{\,Me\kern -0.1em V}}\xspace}
\newcommand{\kev}{\ensuremath{\mathrm{\,ke\kern -0.1em V}}\xspace}
\newcommand{\ev}{\ensuremath{\mathrm{\,e\kern -0.1em V}}\xspace}
\newcommand{\gevc}{\ensuremath{{\mathrm{\,Ge\kern -0.1em V\!/}c}}\xspace}
\newcommand{\mevc}{\ensuremath{{\mathrm{\,Me\kern -0.1em V\!/}c}}\xspace}
\newcommand{\gevcc}{\ensuremath{{\mathrm{\,Ge\kern -0.1em V\!/}c^2}}\xspace}
\newcommand{\gevgevcccc}{\ensuremath{{\mathrm{\,Ge\kern -0.1em V^2\!/}c^4}}\xspace}
\newcommand{\mevcc}{\ensuremath{{\mathrm{\,Me\kern -0.1em V\!/}c^2}}\xspace}

\def\invfb   {\ensuremath{\mbox{\,fb}^{-1}}\xspace}

\def\ps   {\ensuremath{{\mathrm{ \,ps}}}\xspace}

\def\invps{\ensuremath{{\mathrm{ \,ps^{-1}}}}\xspace}

\def\gsim{{~\raise.15em\hbox{$>$}\kern-.85em
          \lower.35em\hbox{$\sim$}~}\xspace}
\def\lsim{{~\raise.15em\hbox{$<$}\kern-.85em
          \lower.35em\hbox{$\sim$}~}\xspace}

\def\rad{\ensuremath{\mathrm{ \,rad}}\xspace}

\def\tell1  {TELL1\xspace}
\def\ukl1   {UKL1\xspace}

\title{Measurements of $\phi_s$ at the LHCb experiment}

\ShortTitle{Measurements of $\phi_s$ at LHCb}

\author{\speaker{Greig A. Cowan}\thanks{On behalf of the LHCb collaboration}\\
        University of Edinburgh, UK\\
        E-mail: \email{g.cowan@ed.ac.uk}}

\abstract{These proceedings present the current status of measurements of the
\CP-violating phase \phis\ by the LHCb collaboration, reviewing the measurements
in channels such as $\Bs\to\jpsi\phi$, $\Bs\to\jpsi\pi^+\pi^-$ and $\Bs\to\psi(2S)\phi$.
The observation of the
$\Bs\to\eta_c\phi$ decay mode is presented for the first time, which can be used 
to measure \phis\ with larger data samples that will be collected over the coming years
by the LHCb experiment.
Finally, the expected increase in precision from LHCb measurements of \phis\ over the
next decade is presented.
}

\FullConference{9th International Workshop on the CKM Unitarity Triangle\\
		28  November - 3 December 2016\\
		Tata Institute for Fundamental Research (TIFR), Mumbai, India}

\usepackage{cite}
\usepackage{mciteplus}
\usepackage{verbatim}
\begin{document}

\section{Introduction and motivation}

A key observable to be measured in the \Bs meson system is
the \CP-violating phase, \phis, which arises due to the interference
between \Bs meson mixing and decay processes.
It is defined as $\phi_{s} \equiv -{\rm arg}(\lambda_f) \equiv
-{\rm arg}\left(\frac{q}{p}\frac{A_f}{\overline{A}_f}\right)$, where $q, p$ are
complex eigenvalues related to \Bs mixing and $A_f$ ($\overline{A}_f$) are the complex
amplitudes for \Bs (\Bsb) meson decay to final state $f$.
Global fits to experimental data give a precise prediction
for $\phi_s$ in the Standard Model of $-0.0376\pm 0.0008\rad$~\cite{Charles:2015gya}.
Any deviation from this prediction would be a clear sign for non-Standard Model
physics, strongly motivating the need for precise experimental measurements of 
this quantity.
In this article I will review the measurements of this observable from the LHCb
collaboration and discuss new measurements of \Bs meson decay channels that
can be used to measure \phis\ in the future. All measurements shown here
use 3\invfb of data collected by the LHCb experiment~\cite{Alves:2008zz} in $pp$ collisions
at the LHC during 2011 and 2012.

\section{State-of-the-art of $\phi_s$ measurements}
\subsection{$\phi_s$ from $\Bs\to \jpsi\phi$ and $\Bs\to \jpsi\pi^+\pi^-$}

\begin{figure}[t]
\centering
\includegraphics[width=0.24\linewidth]{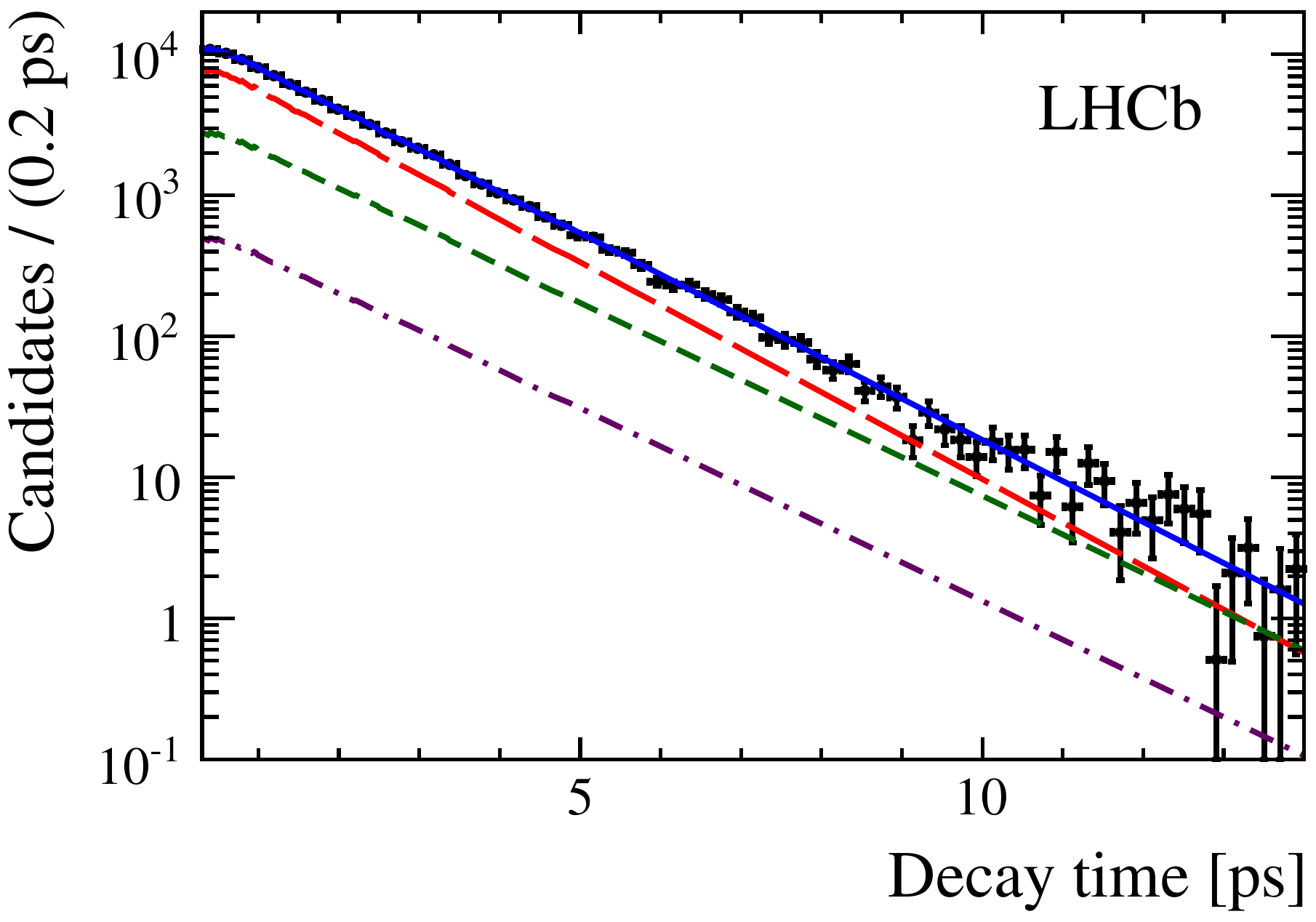}
\includegraphics[width=0.24\linewidth]{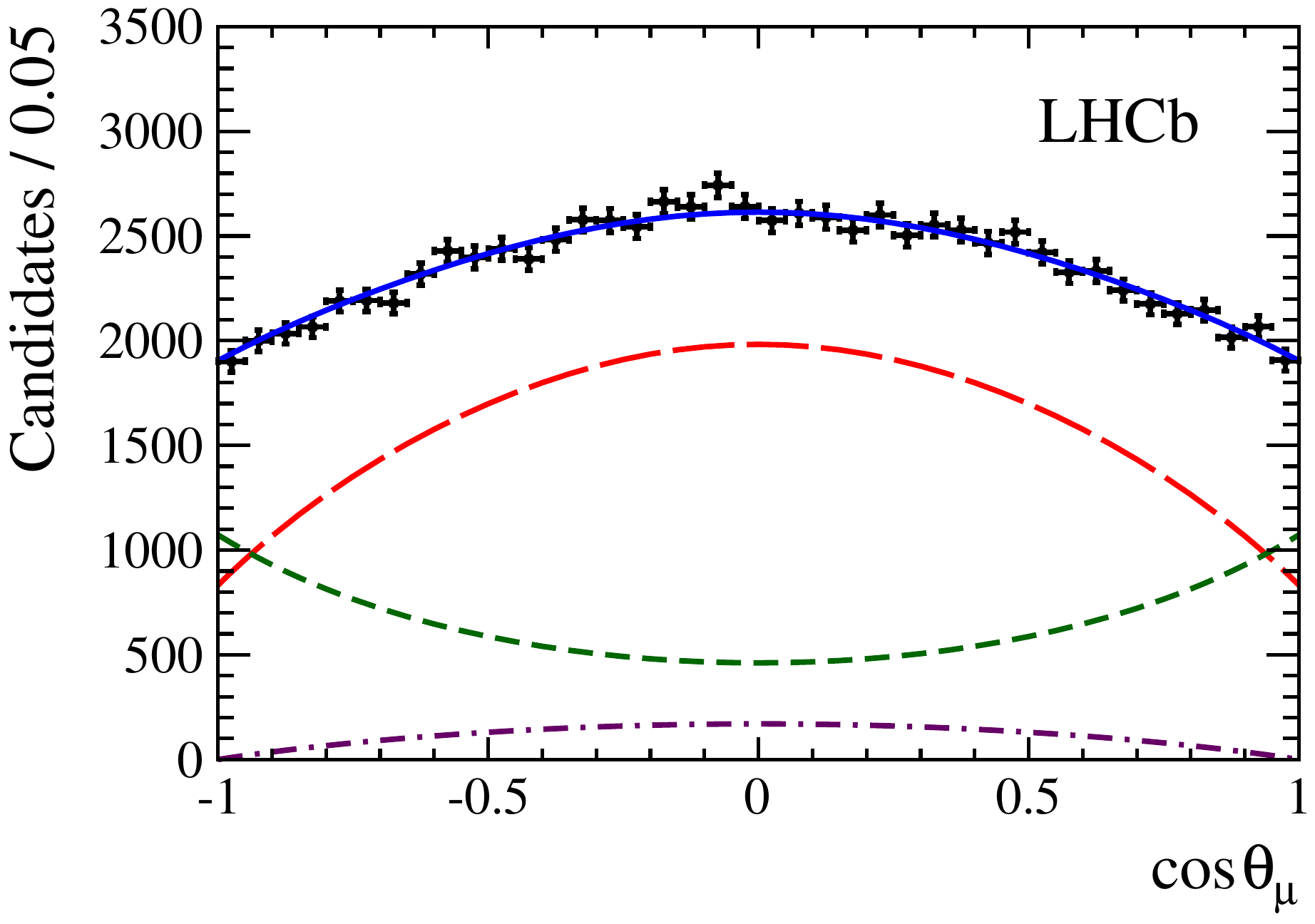}
\includegraphics[width=0.24\linewidth]{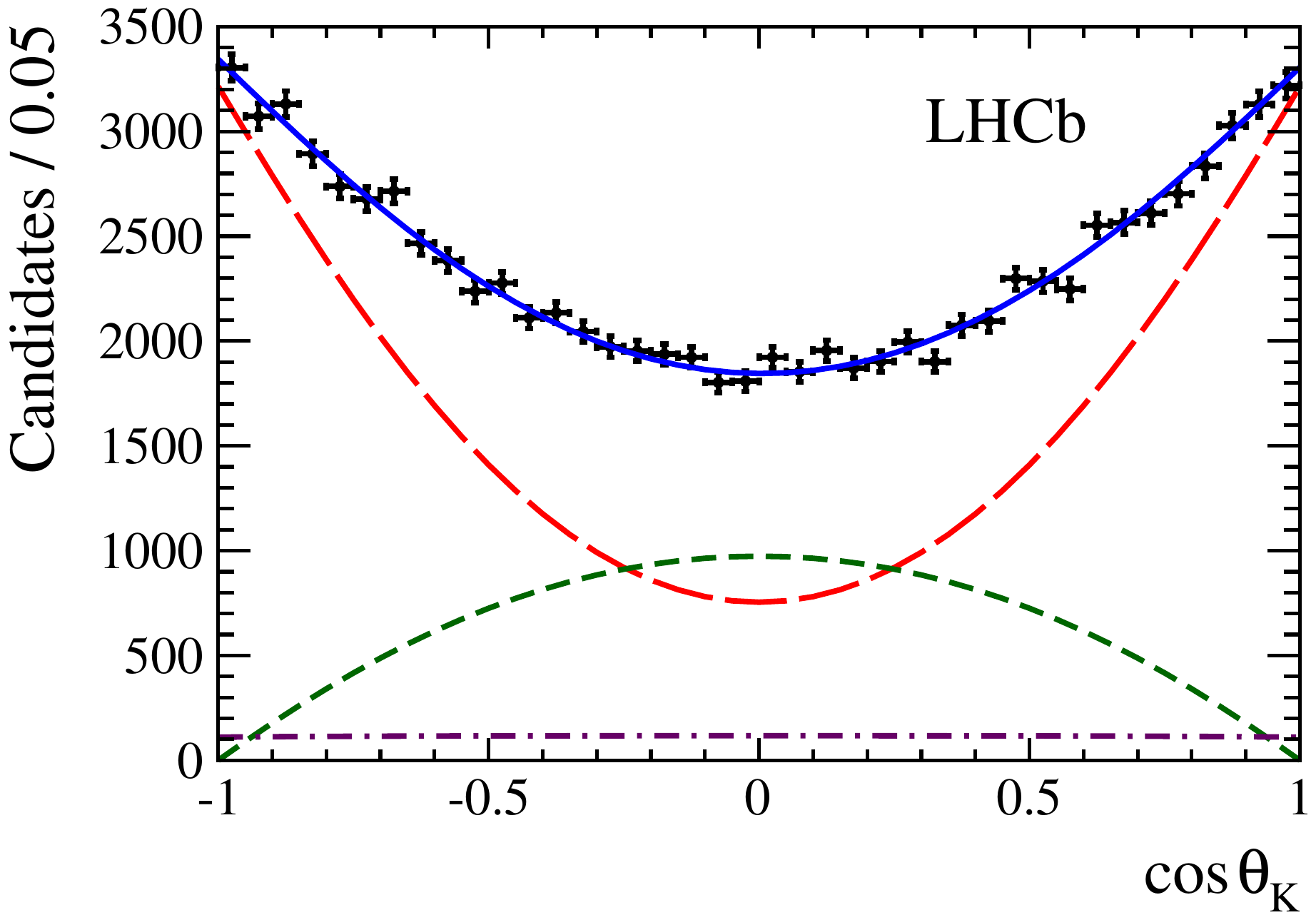}
\includegraphics[width=0.24\linewidth]{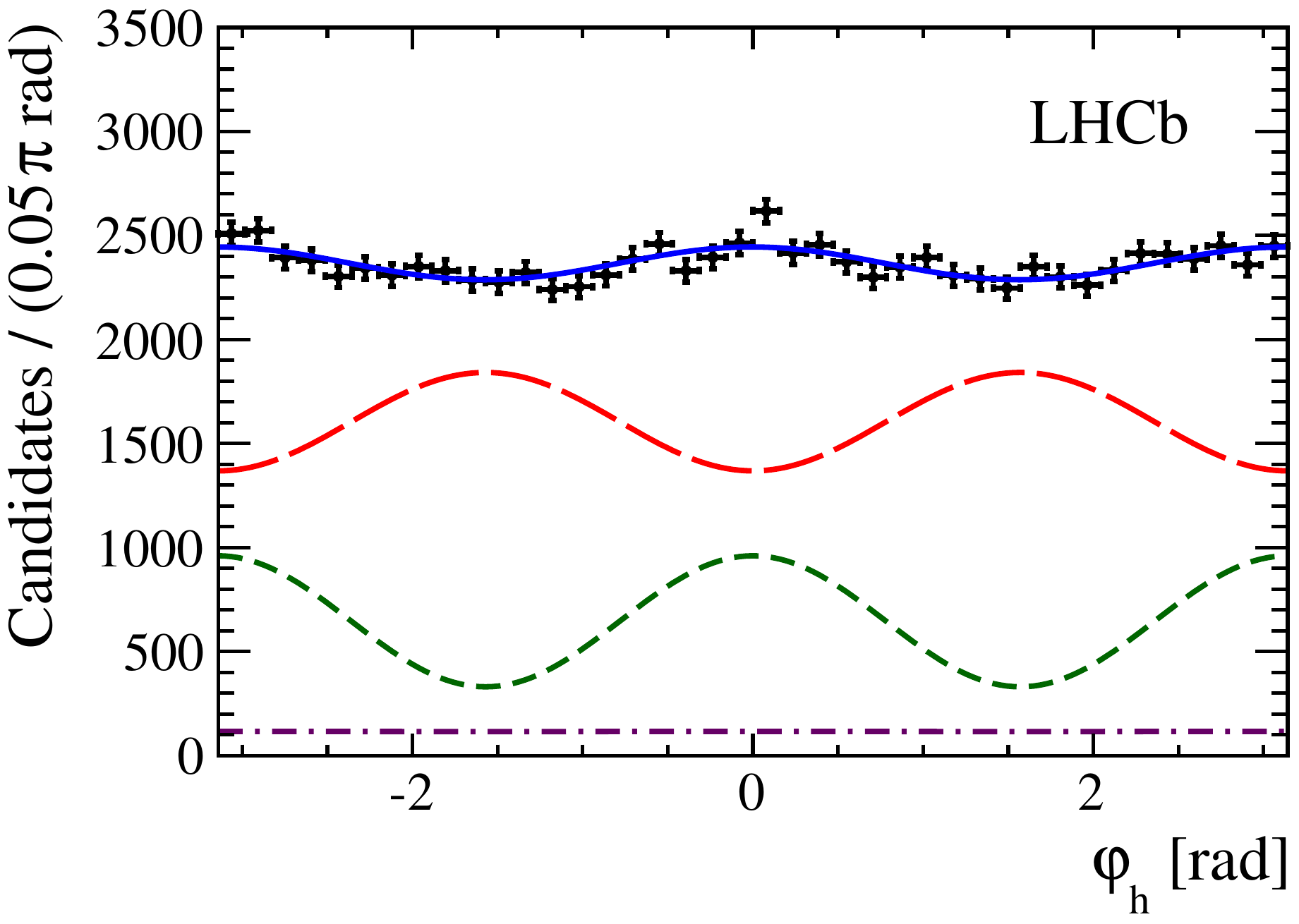}
\caption{\label{fig:Bs2JpsiPhi}\small  Projections of $\Bs\to \jpsi\phi$ data onto the 
decay time and three helicity angles. The projection of the total fit model is shown (blue)
along with the individual \CP-odd (green), \CP-even (red) and S-wave (purple) contributions.}
\end{figure}

The so-called ``golden mode" for measuring \phis\ is using a flavour-tagged
time-dependent angular analysis of the $\Bs\to \jpsi\phi$ decay, where
$\jpsi\to\mu^+\mu^-$ and $\phi\to\Kp\Km$. This $b\to c\overline{c}s$
mediated decay has a high branching fraction
and the presence of two muons in the final state leads to a high trigger
efficiency. The angular analysis is necessary to disentangle the interfering
\CP-odd and \CP-even components in the final state, which arise due to the
relative angular momentum between the two vector resonances.
In addition, there is a small ($\sim 2\%$) \CP-odd $\Kp\Km$ S-wave
contribution that must be accounted for.
The LHCb detector has excellent time resolution ($\sim 45$ fs~\cite{LHCb-PAPER-2013-002})
and tagging power ($\sim 4\%$~\cite{LHCb-PAPER-2015-056}), both of which are crucial to the measurement.
In Run 1, the LHCb collaboration used a sample of $\sim 96000$  
$\Bs\to \jpsi\phi$ decays to measure \phis, the width difference
between the light and heavy \Bs mass eigensates ($\Delta\Gamma_s$),
the average decay time ($\Gamma_s$), mixing frequency ($\Delta m_s$) 
and direct \CP violation parameter ($|\lambda|$).
Figure~\ref{fig:Bs2JpsiPhi} shows the results of this analysis, which gave
 $\phis = -0.058   \pm 0.049              \pm  0.006$~rad, 
 $\DGs = 0.0805  \pm 0.0091         \pm  0.0032$~ps$^{-1}$ and
 $\Gamma_s = 0.6603 \pm 0.0027        \pm  0.0015$~ps$^{-1}$ 
 ~\cite{LHCb-PAPER-2014-059}.
 These are the most precise determinations of these parameters to date
 and are consistent with  SM predictions~\cite{Charles:2015gya,Artuso:2015swg}.
The dominant systematic uncertainties in these measurement arise from knowledge
about the decay time and angular efficiencies.

It is possible that due to unknown hadronic effects or beyond the SM physics, the
values of \phis\ and $|\lambda|$ could be different for each of the four polarisation
states~\cite{Faller:2008gt,Bhattacharya:2012ph}. For the first time, the LHCb collaboration
relaxed this assumption in the analysis, finding that no polarisation dependence was visible within the available
statistical precision.

The LHCb collaboration has also used a similar analysis of $\Bs\to \jpsi\pi^+\pi^-$
decays to measure \phis~\cite{LHCb-PAPER-2014-019}.
Here, the full $\pi^+\pi^-$ mass spectrum is used in the measurement, 
which has previously been studied and found to be
$>97.7\%$ completely \CP-odd~\cite{LHCb-PAPER-2012-005}, dominated by the $f_0(980)$
component. With this time-dependent amplitude analysis, $\phi_s$ was measured to be
$0.070  \pm  0.068  \pm  0.008$~rad, 
the dominant systematic uncertainty coming from knowledge about the composition of resonances in the 
$\pi^+\pi^-$ spectrum. Since the final state is almost all \CP-odd, a simplified tagged fit to only
the \Bs decay time distribution yields compatible results. Combining the
$\Bs\to \jpsi\phi$ and $\Bs\to \jpsi\pi^+\pi^-$ results gives $\phis = -0.010  \pm  0.039$~rad.

\subsection{$\phi_s$ from $\Bs\to \psi(2S)\phi$}

\begin{figure}[t]
\centering
\includegraphics[width=0.31\linewidth]{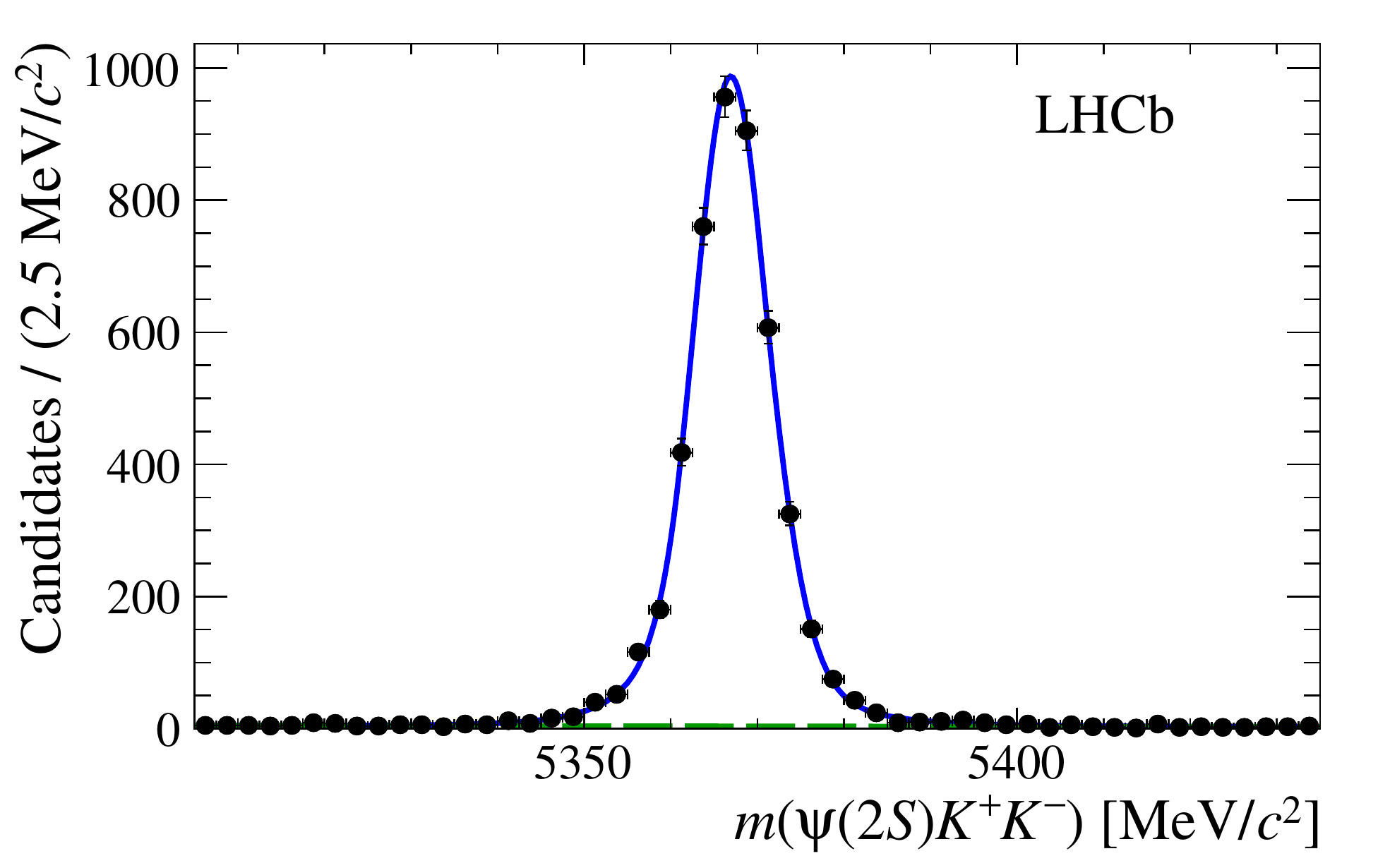}
\includegraphics[width=0.3\linewidth]{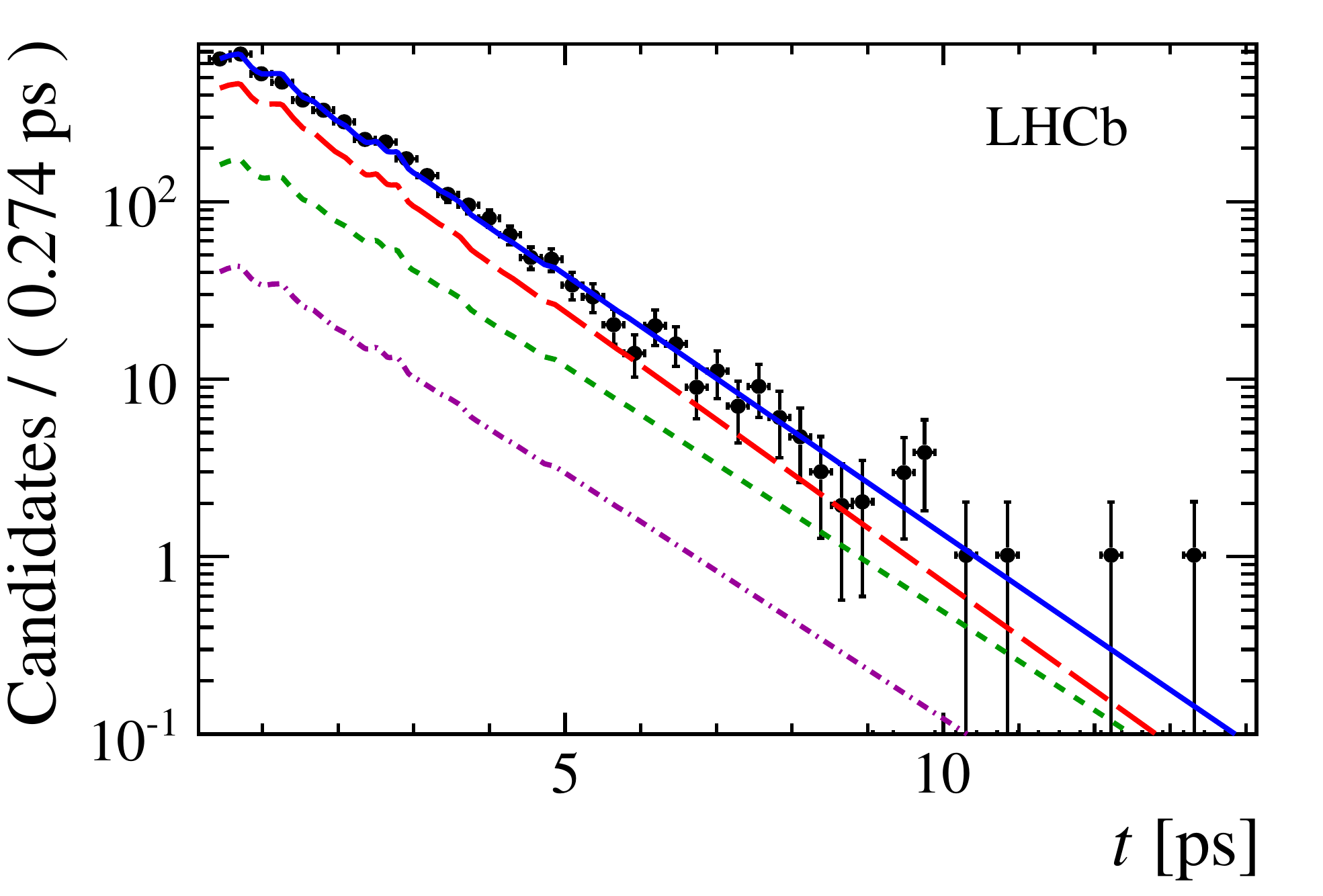}
\includegraphics[width=0.3\linewidth]{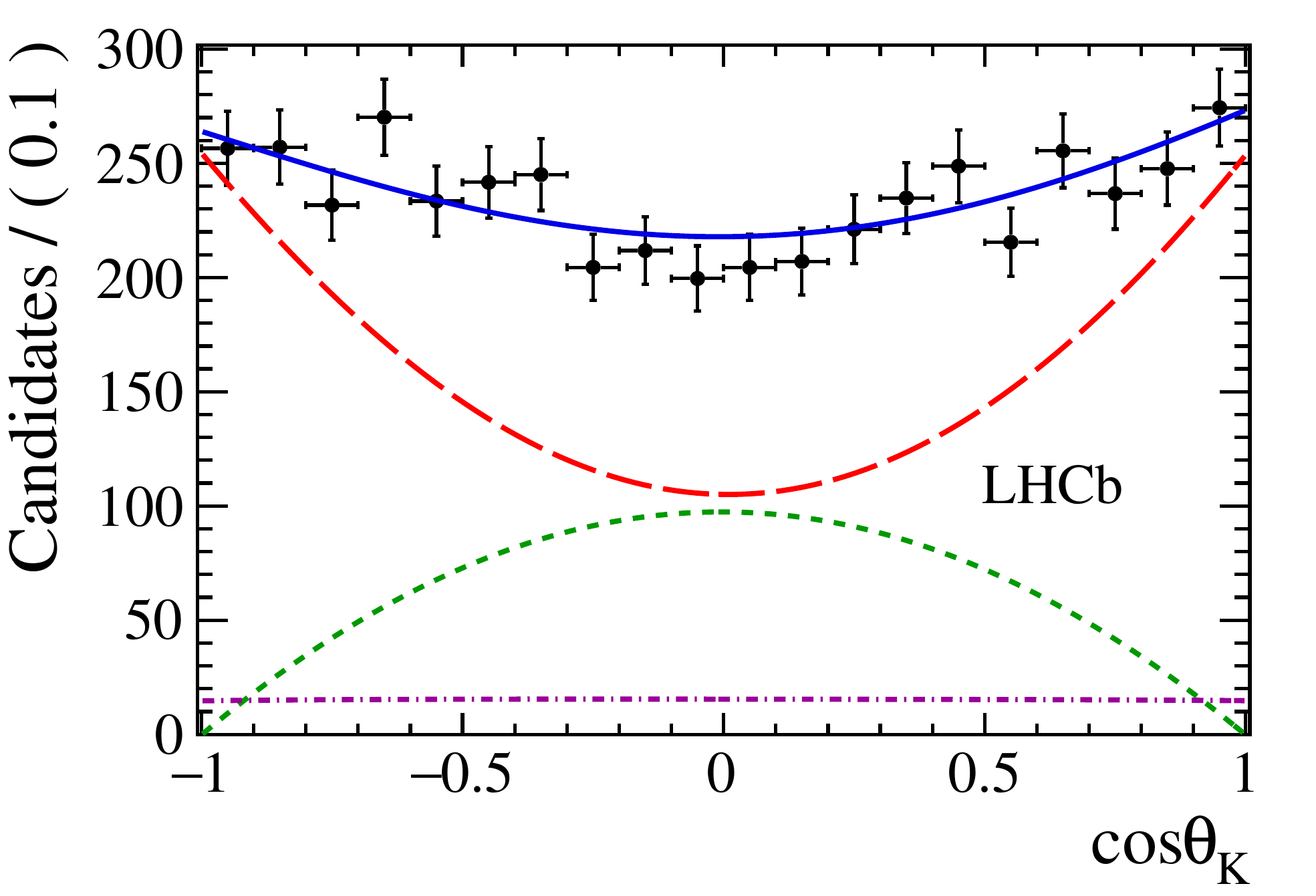}\\
\hspace{4.7cm}
\includegraphics[width=0.3\linewidth]{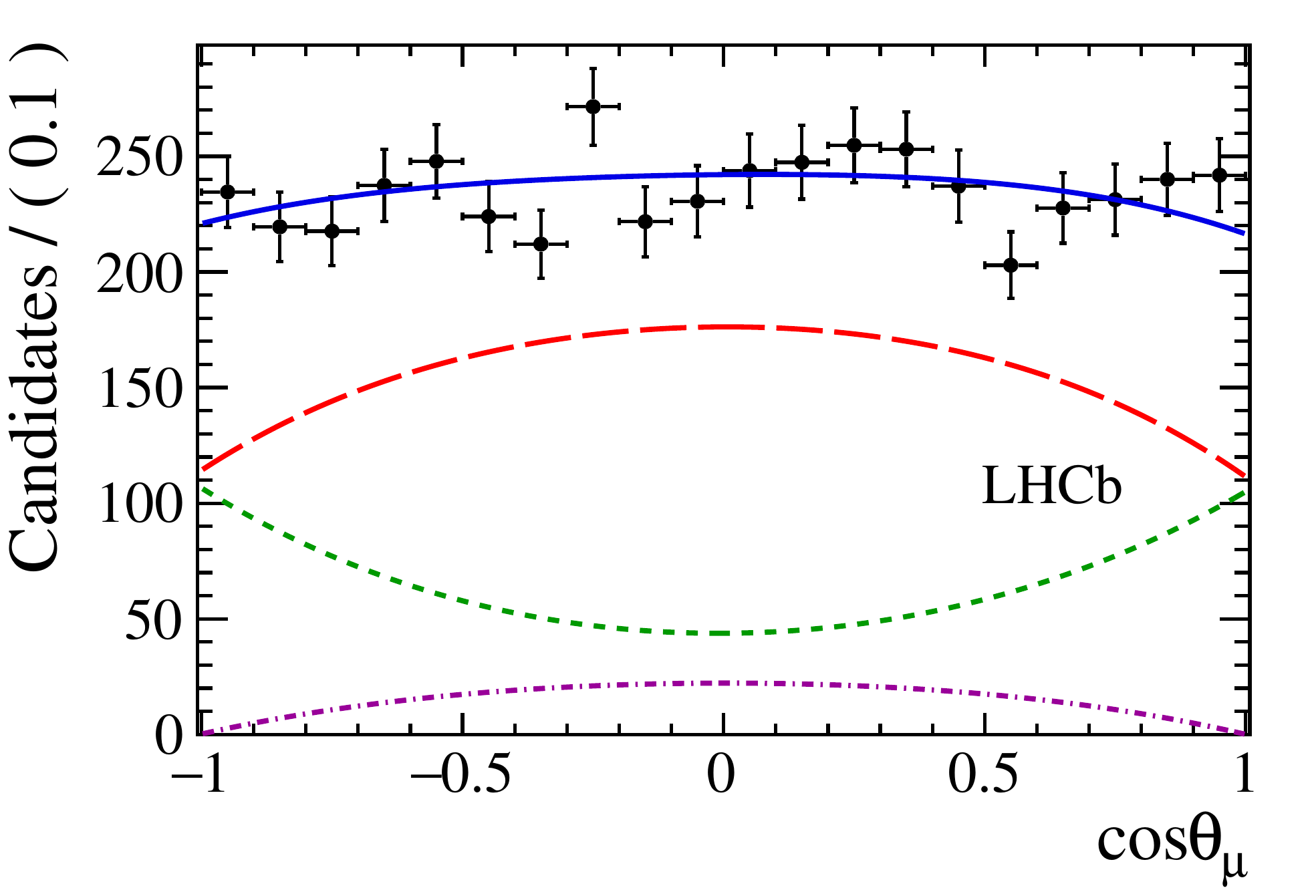}
\includegraphics[width=0.3\linewidth]{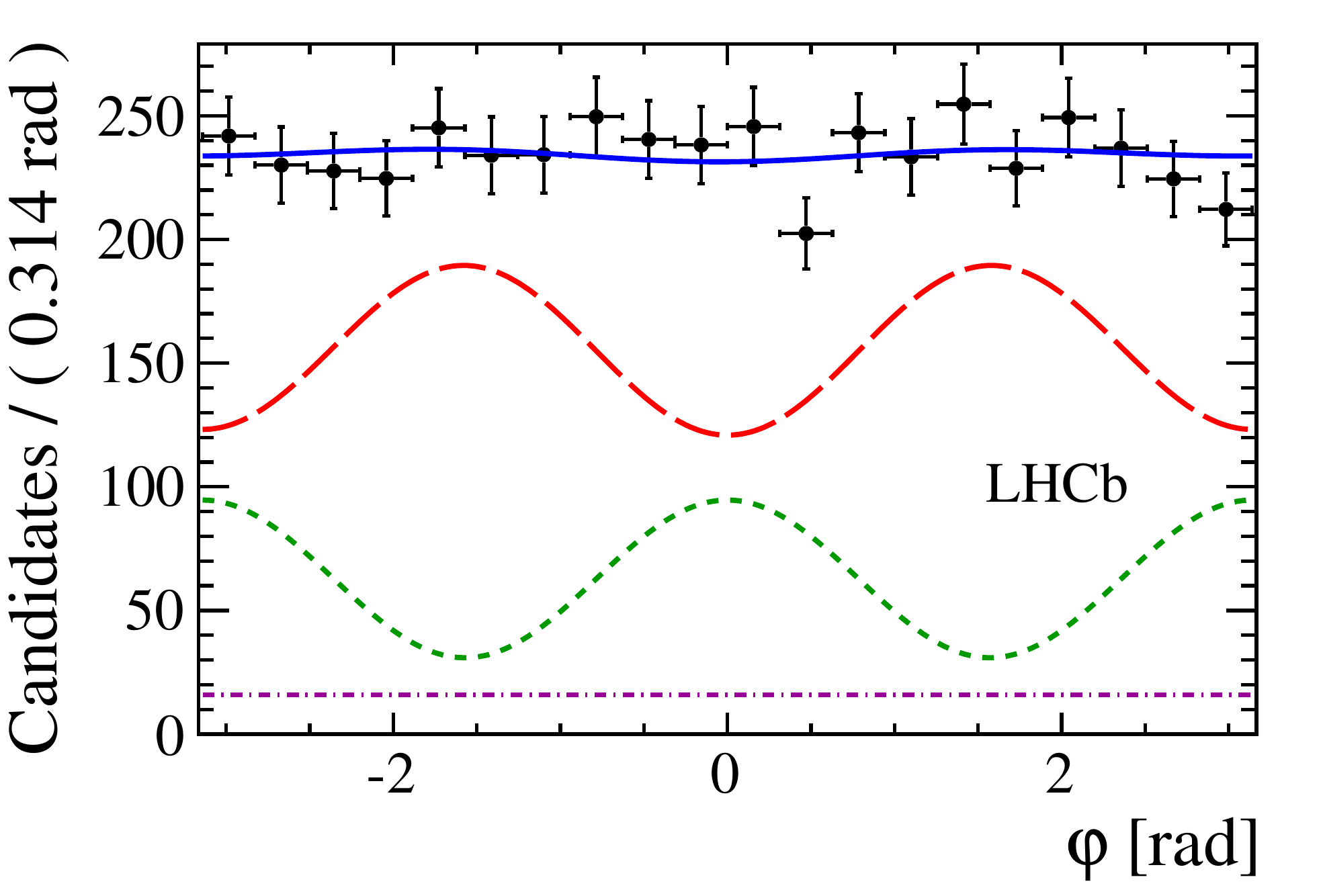}
\caption{\label{fig:Bs2psi2SPhi}\small  Distribution of $\psi(2S)\phi$ invariant mass of selected
$\Bs\to \psi(2S)\phi$ candidates and projections of $\Bs\to \psi(2S)\phi$ data and fit model
(see Figure 1 for legend).}
\end{figure}

Other \Bs decay modes with $b\to c\overline{c}s$ transitions can be used to 
measure \phis. In Ref.~\cite{LHCb-PAPER-2016-027}, LHCb studied the
$\Bs\to \psi(2S)\phi$ (with $\psi(2S)\to\mu^+\mu^-$) decays for the first time
using the same analysis techniques as Ref.~\cite{LHCb-PAPER-2014-059}.
Figure~\ref{fig:Bs2psi2SPhi} shows  $\sim 4500$ signal decays in Run 1 data, 
selected using a boosted decision tree that has been trained using simulated
signal events and a background sample from the high-mass sideband.
Figure~\ref{fig:Bs2psi2SPhi} also shows the projections of the data and fit
onto the decay time and helicity angles, demonstrating a good fit to the data.
In addition to $\Delta\Gamma_s$ and $\Gamma_s$, $\phis$ was measured
to be $0.23^{+0.29}_{-0.28} \pm 0.02$ rad.
For the first time the magnitude of the transversity amplitudes and their phases
were measured for this decay, which are different to those in $\Bs\to \jpsi\phi$
as expected~\cite{Hiller:2013cza}.

\subsection{Global combination}

\begin{figure}[t]
\centering
\includegraphics[width=0.4\linewidth]{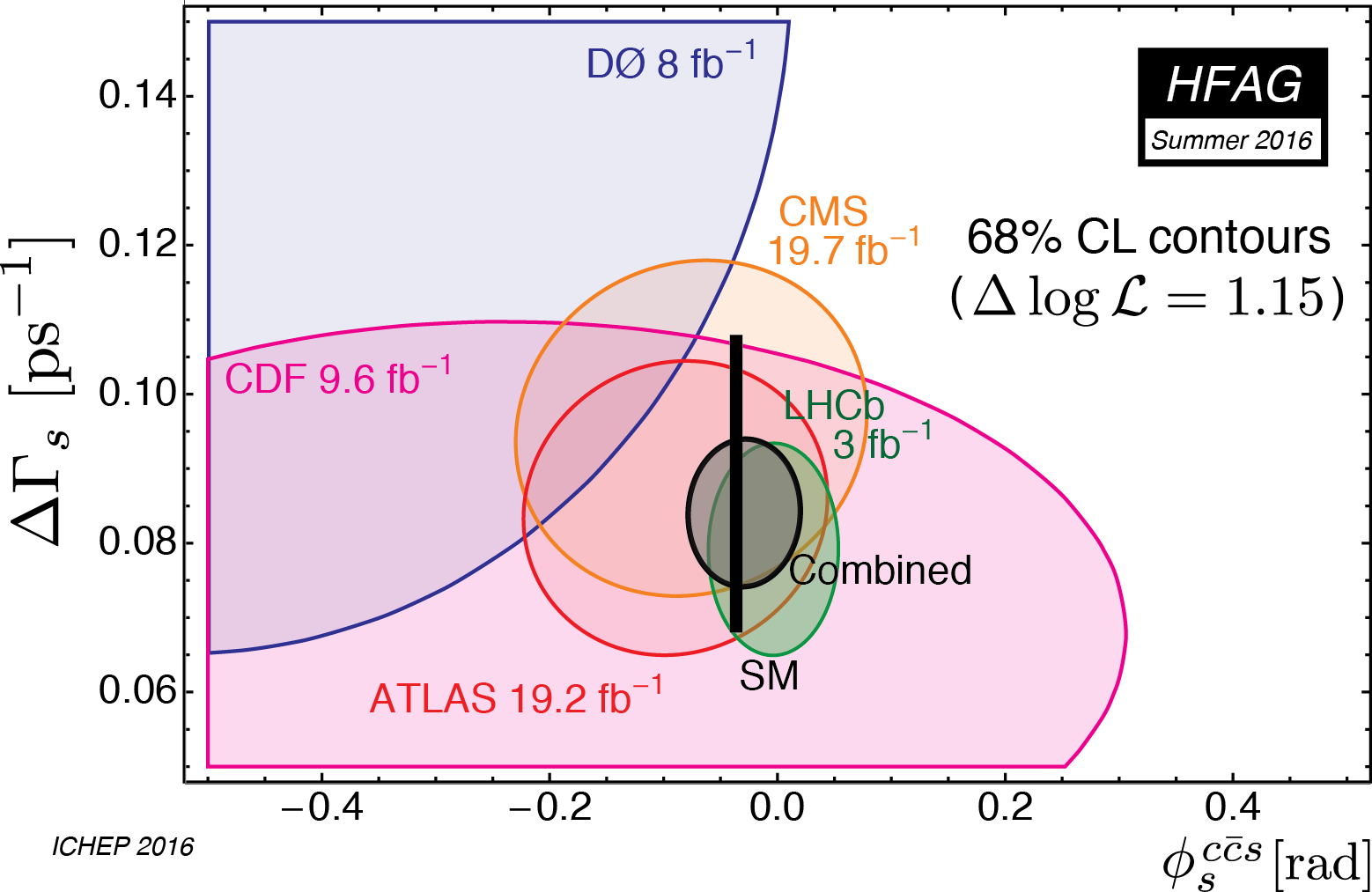}
\caption{\label{fig:comb}\small  HFAG combination~\cite{Amhis:2016xyh} of \phis\ and $\Delta\Gamma_s$ from several
experiments.}
\end{figure}

The global combination of \phis\ and $\Delta\Gamma_s$ measurements from the Heavy Flavour Averaging Group~\cite{Amhis:2016xyh} is shown in Figure~\ref{fig:comb}, using measurements
from the LHCb collaboration discussed here along with those
from the CDF~\cite{Aaltonen:2012ie}, D0~\cite{Abazov:2011ry},
ATLAS~\cite{Aad:2016tdj} and CMS~\cite{Khachatryan:2015nza} collaborations. They find
$\DGs    =  0.085   \pm  0.006 \invps$ and $\phi_s  =  -0.030  \pm  0.033$~rad.
The results are dominated by 
those from the LHCb collaboration and are consistent with the SM predictions. There remains
space for new physics contributions at the $\sim 20\%$ level, however, as the experimental 
precision improves, it is essential that there is good control over hadronic effects (so-called
``penguin pollution") that could mimic the effect from beyond-the-SM physics.

\subsection{$\phi_s^{ss\overline{s}}$ from $\Bs\to \phi\phi$}

A related \CP-violating phase, $\phi_s^{ss\overline{s}}$, can be
measured by applying similar methods as above to \Bs meson decays that
go via a $b \to ss\overline{s}$ transition. The LHCb collaboration has 
performed such an analysis using $\Bs\to \phi\phi$~\cite{LHCb-PAPER-2014-026},
measuring $\phis = -0.17 \pm 0.15 \pm 0.03$ rad, which is consistent with the Standard Model
predictions, all of which are very close to zero~\cite{Beneke:2006hg,Bartsch:2008ps,Cheng:2009mu}.
An upcoming study of $\Bs\to \Kp\pi^-\Kp\pi^-$ decays will provide another
avenue for measuring this quantity~\cite{Garcia}.

\section{Future prospects for measuring $\phi_s$}

The measurement of \phis\ using $\Bs\to \jpsi\phi$ decays has so far restricted 
to using the region of $\Kp\Km$ phase space near the $\phi(1020)$ resonance. 
A full amplitude analysis of the  $\Bs\to \jpsi\Kp\Km$ system was performed
in Ref.~\cite{LHCb-PAPER-2016-028}, indicating a significant contribution from
other $\Kp\Km$ resonances such as the $f_2^\prime(1525)$ that can be used
when measuring \phis\ to increase the statistical precision.
This approach will require the application of the same analysis formalism as in Ref~\cite{LHCb-PAPER-2014-019}. 
Similarly, the recently observed $\Bs\to \phi\pi^+\pi^-$ decay~\cite{LHCb-PAPER-2016-028}
could be used with 
future data samples from Run 2 and beyond
to measure $\phi_s^{ss\overline{s}}$, again with a flavour-tagged, decay-time dependent
amplitude analysis, including all appropriate $\pi^+\pi^-$ resonances.

\subsection{Observation of $\Bs\to\eta_c\phi$}

\begin{figure}[t]
\centering
\includegraphics[width=0.39\linewidth]{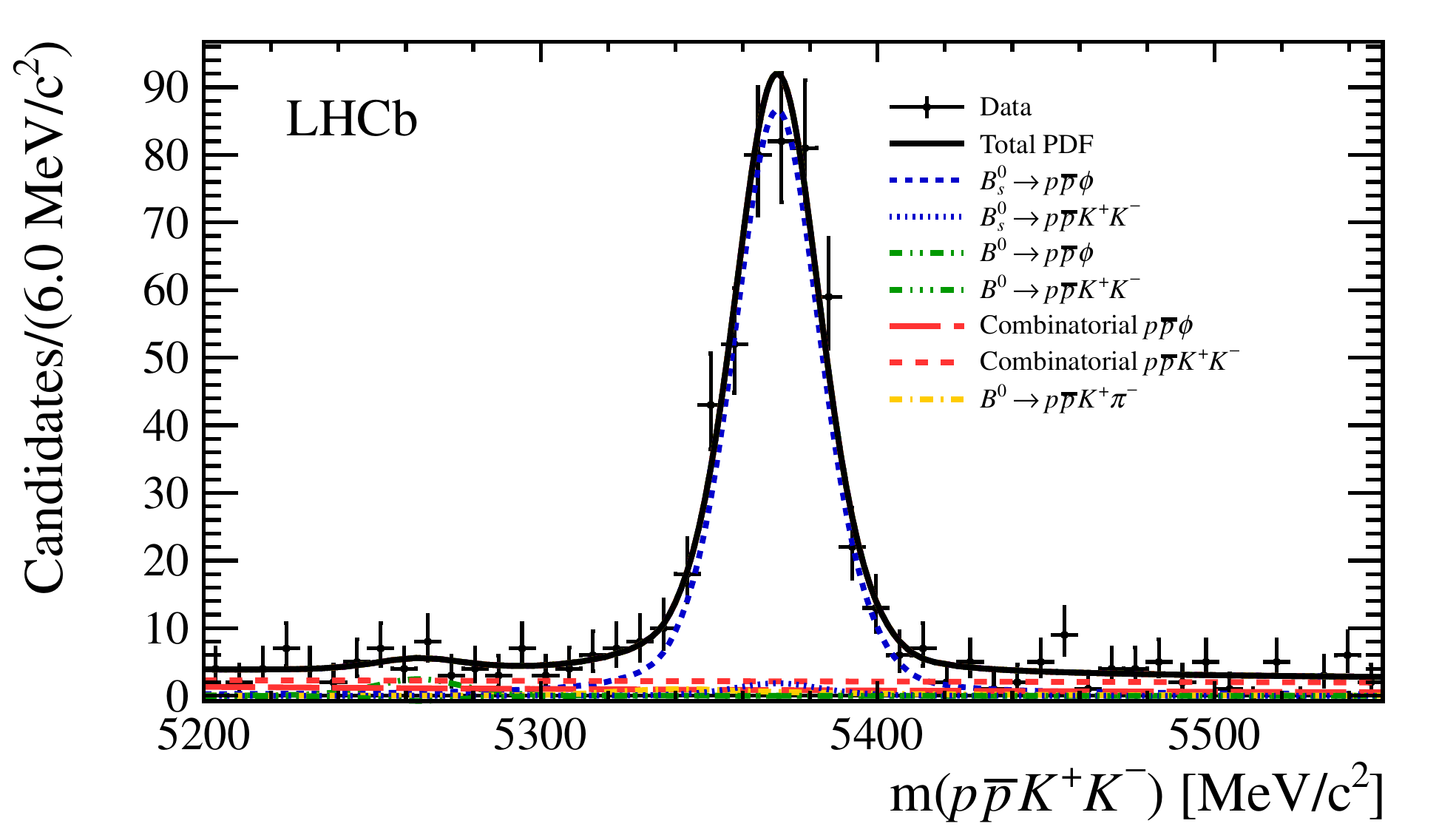}
\includegraphics[width=0.33\linewidth]{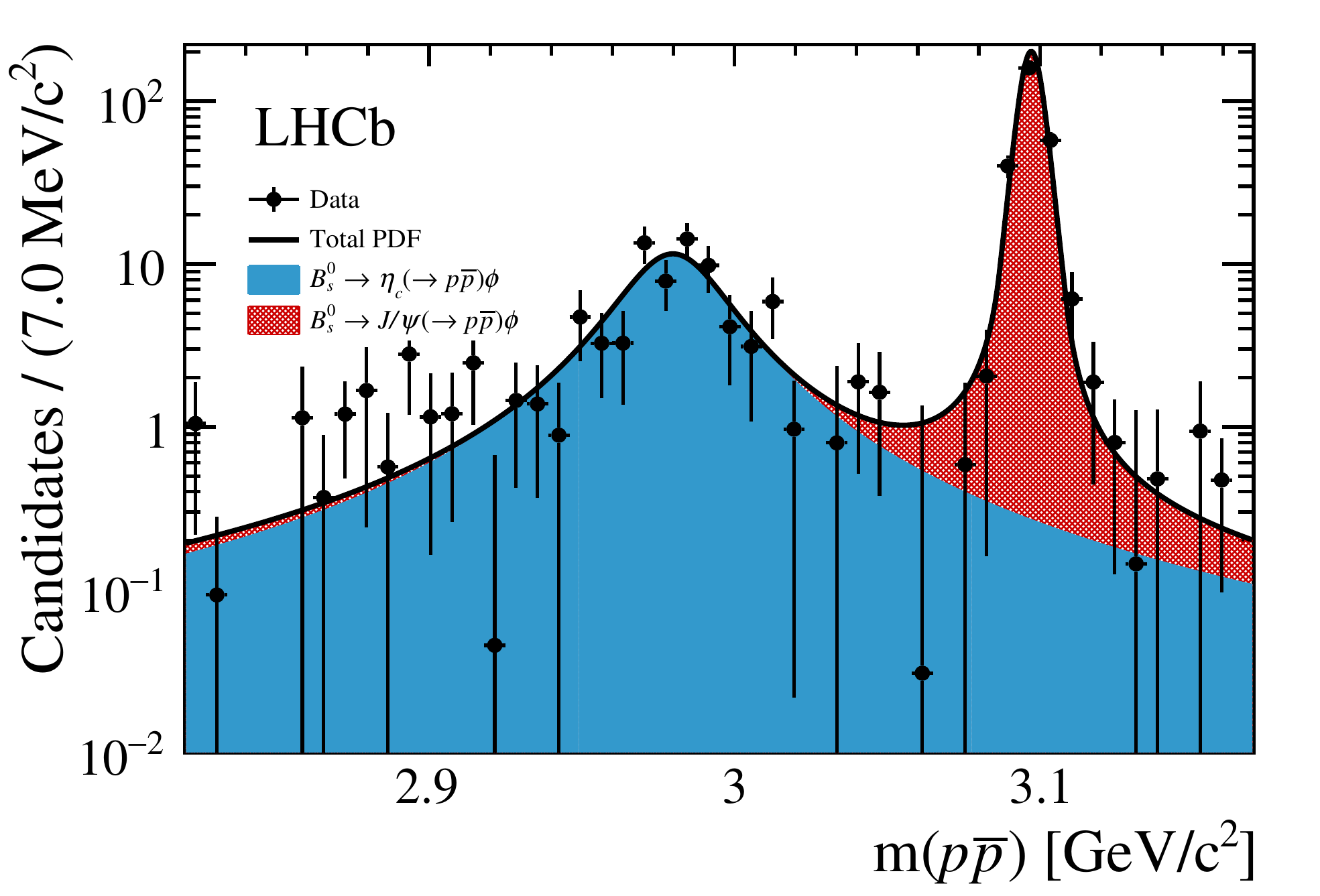}
\caption{\label{fig:eta_c_phi}\small Invariant mass distributions for selected $\Kp\Km p\overline{p}$ (left)
and $p\overline{p}$ (right) candidates.}
\end{figure}

At this conference the LHCb collaboration announced a preliminary observation
of the $\Bs\to\eta_c\phi$ decay mode, with 
$\eta_c\to \Kp\Km\pi^+\pi^-, \pi^+\pi^-\pi^+\pi^-$, $\Kp\Km\Kp\Km, p\overline{p}$~\cite{LHCb-PAPER-2016-056}.
This decay is another $b\to c\overline{c}s$ transition that could be used to measure \phis.
Figure~\ref{fig:eta_c_phi} shows the invariant mass of the $\Bs$ system in the 
$p\overline{p}$ mode along with the $p\overline{p}$ spectrum, with the 
$\eta_c$ and \jpsi charmonium resonances clearly visible. A simultaneous amplitude fit is performed using
all modes and including contributions from interfering non-resonant components. The
branching fraction is extracted relative to the \jpsi\ channel and found to be
${\cal B}(\Bs\to\eta_c\phi) = (5.01 \pm 0.53 ({\rm stat}) \pm 0.27 ({\rm syst}) \pm 0.63 ({\cal B}) ) \times 10^{-4}$.
First evidence of the $\Bs\to\eta_c\pi^+\pi^-$ decay was also presented.

\subsection{$\Bs \to \jpsi \eta$ effective lifetime}

\begin{figure}[t]
\centering
\includegraphics[width=0.35\linewidth]{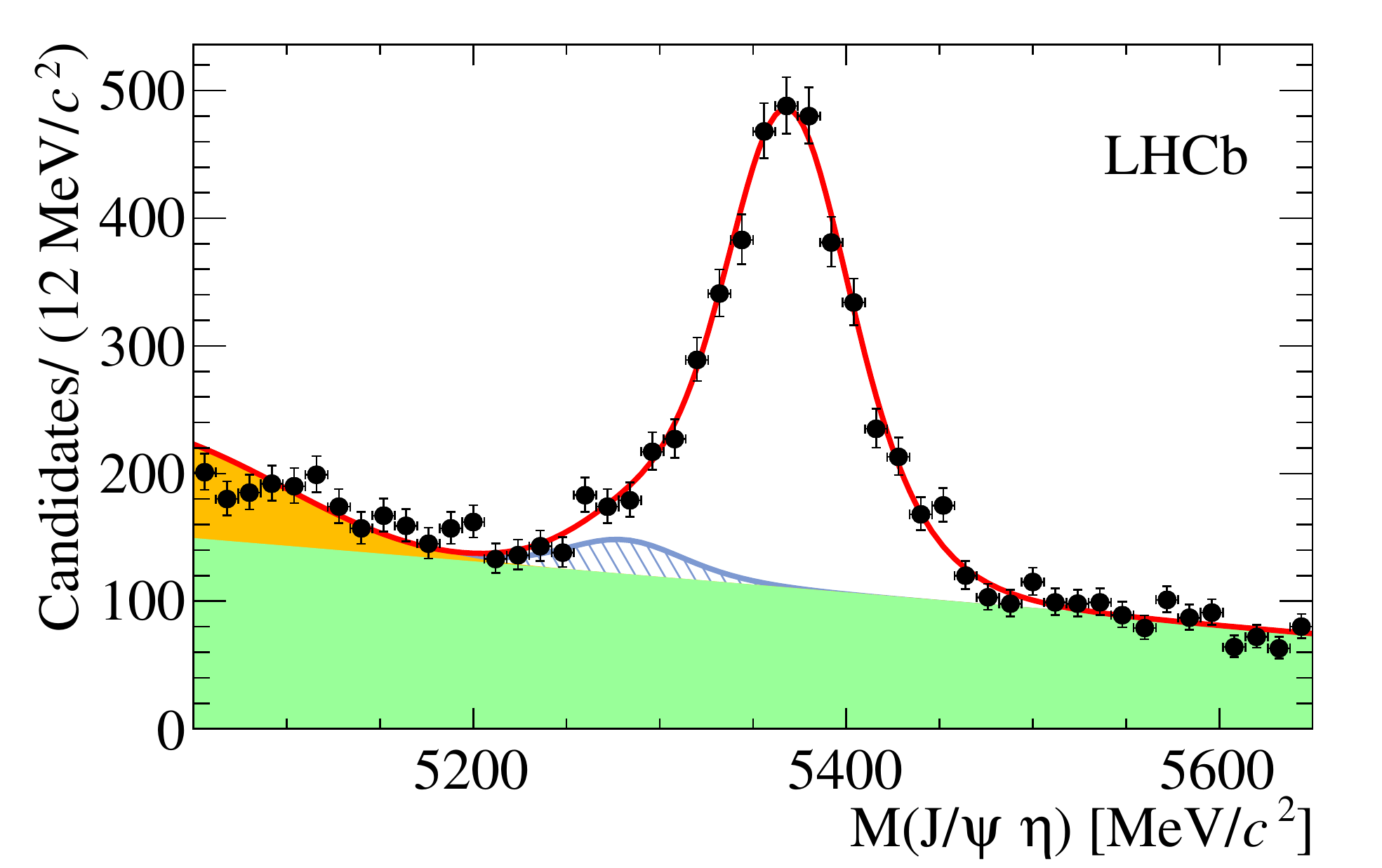}
\includegraphics[width=0.35\linewidth]{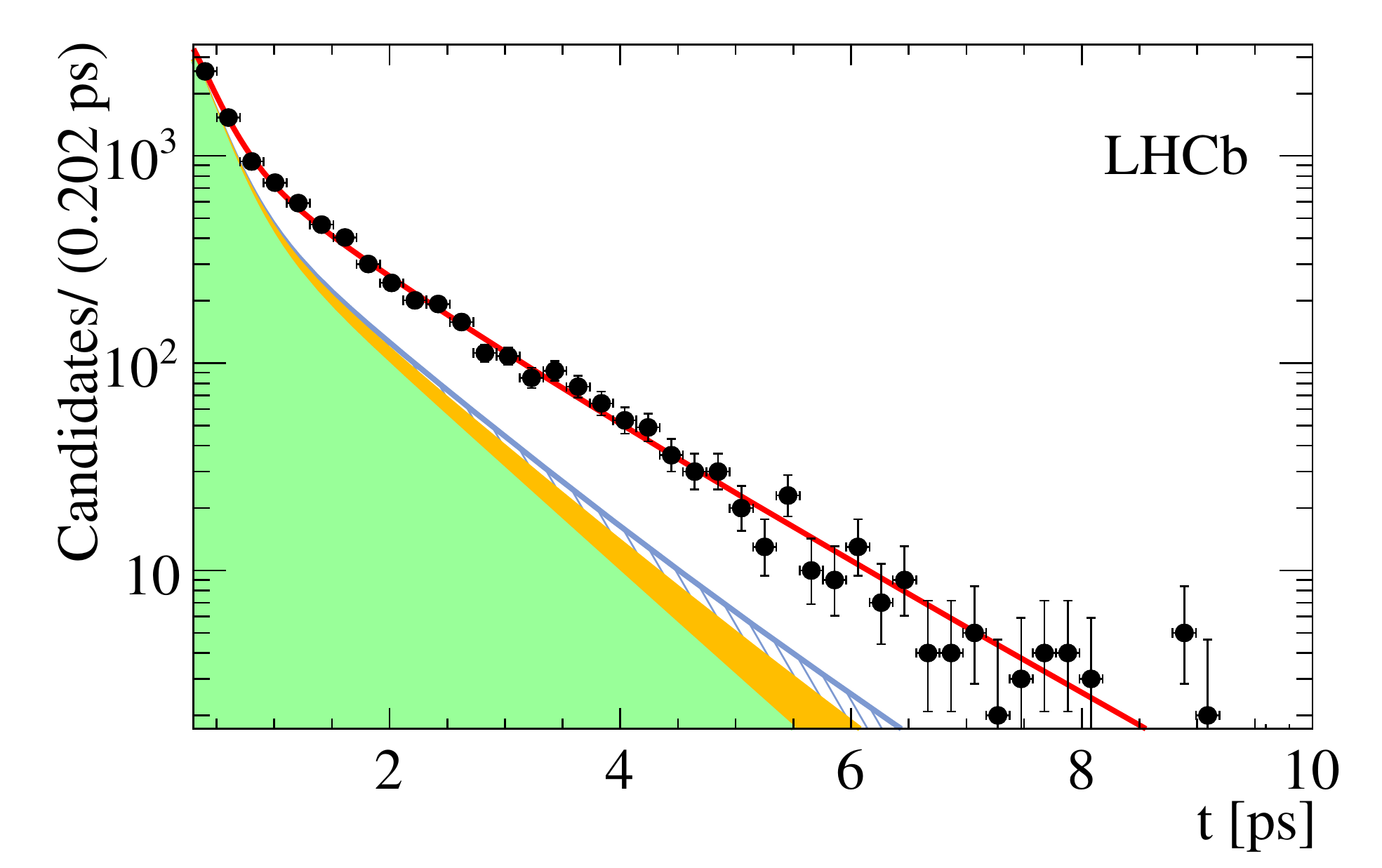}
\caption{\label{fig:lifetime}\small  Distributions of $\jpsi\eta$ invariant mass and decay time for
selected $\Bs\to\jpsi\eta$ decays. Combinatorial background (green), partially reconstructed background (orange)
and background from $\Bd\to\jpsi\eta$ decays (blue) are shown.}
\end{figure}

The LHCb collaboration has recently observed the $\Bs \to \jpsi \eta(\to\gamma\gamma)$
decay~\cite{LHCb-PAPER-2016-017} and used it to measure the \Bs effective
lifetime. As this mode is a \CP-even eigenstate the effective lifetime gives a  measurement
of $\Gamma_{\rm L}$. The final state is challenging, containing only two charged tracks
and the invariant mass resolution is $\sim 48\mevcc$ (see Figure~\ref{fig:lifetime}),
compared to $\sim 8\mevcc$ for $\Bs\to\jpsi\phi$ decays.
Using $\sim 3000$ signal candidates,
the lifetime was measured to be $\tau = 1.479 \pm 0.034 \pm 0.011$\ps, consistent
with other measurements of the \CP-even lifetime~\cite{LHCb-PAPER-2013-060,LHCb-PAPER-2014-011}.
In the future the $\Bs \to \jpsi \eta$ mode can be used to measure \phis\ from a
flavour-tagged fit to the decay time distribution.

An update of the HFAG
averages of $\Delta\Gamma_s$ and $\Gamma_s$ was presented, showing good 
consistency between all measurements and the SM predicitions~\cite{Artuso:2015swg}.
The $\Delta\Gamma_s$ prediction has
an uncertainty more than three times larger than the experimental average.

\section{Summary}

\begin{figure}[t]
\centering
\includegraphics[width=0.45\linewidth]{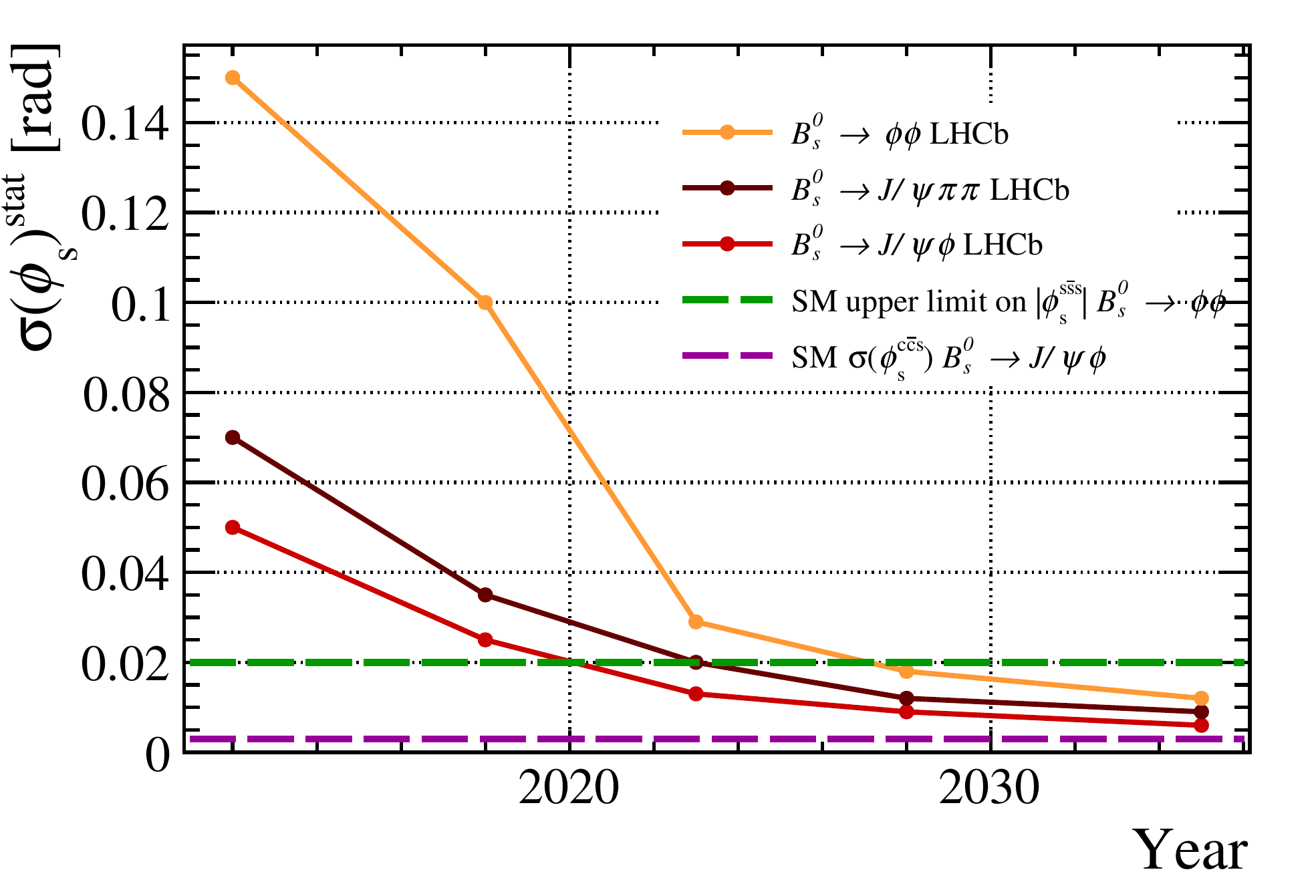}
\caption{\label{fig:projections}\small  Projection of how precision on \phis\ from LHCb
measurements will scale as a function of
time for different decay modes. Information taken from Ref.~\cite{LHCb-PUB-2014-040}.}
\end{figure}

The LHCb collaboration has made leading measurements of the \CP-violating phase $\phi_s$
and \Bs meson lifetimes using Run-1 data. So far all measurements are consistent with 
predictions from the Standard Model. New $b\to c\overline{c}s$ decay modes have
been investigated and measurements performed to either measure \CP violating effects or 
make preparations for such measurements in the future.
Figure~\ref{fig:projections} shows how the precision on \phis\ and $\phi_s^{ss\overline{s}}$
will reduce as a function of time for key decay channels discussed in these proceedings.
The precision is expected to reach
$\sim0.01$\rad at end of Run 3~\cite{LHCb-PUB-2014-040} (the LHCb upgrade era)
which is further discussed in Ref.~\cite{Chob}. As the precision improves it will be
essential to control hadronic effects that can hide small contributions from
non-Standard Model physics~\cite{Akar}.

\section{Acknowledgements}

The author thanks the organisers of the CKM2016 meeting and acknowledges the support of
the Science and Technology Facilities Council (UK) grant ST/K004646/1.

\addcontentsline{toc}{section}{References}
\setboolean{inbibliography}{true}
\bibliographystyle{LHCb}
\bibliography{LHCb-PAPER}

\end{document}